\documentclass[aps,prl,twocolumn,superscriptaddress,longbibliography]{revtex4-2}

\usepackage{amsmath,amssymb}
\usepackage{graphicx}
\usepackage[colorlinks=true,allcolors=blue]{hyperref}
\usepackage{siunitx}
\usepackage{braket}
\usepackage{overpic}
\usepackage{float}
\usepackage{orcidlink}
\usepackage{grffile}

\begin{document}

\title{Spatially Resolved Temperature Measurement Using Rydberg Doppler Broadening Thermometry}

\author{K. N. Trivedi\,\orcidlink{0009-0006-3061-7978}}

\affiliation{Dipartimento di Fisica ‘E. Fermi’, Universit\`a di Pisa, Largo Pontecorvo 3, 56127 Pisa, Italy}
\affiliation{CNR-INO, Via G. Moruzzi 1, 56124 Pisa, Italy}

\author{M. Carminati\,\orcidlink{0009-0002-9232-731X}}
\affiliation{Dipartimento di Fisica ‘E. Fermi’, Universit\`a di Pisa, Largo Pontecorvo 3, 56127 Pisa, Italy}

\author{Èlia Solé Cardona\,\orcidlink{0009-0006-5322-5467}}
\altaffiliation{Present address: University of Florence, Via Nello Carrara, 1, 50019, Sesto Fiorentino, FI, Italy}
\affiliation{CNR-INO, Via G. Moruzzi 1, 56124 Pisa, Italy}

\author{T. Bonaccorsi\,\orcidlink{0000-0001-8666-6142}}
\altaffiliation{Present address: ETH Zürich, Professur für Quanten-Optik, HPF D 20, Otto-Stern-Weg 1, 8093 Zürich, Switzerland}
\affiliation{Dipartimento di Fisica ‘E. Fermi’, Universit\`a di Pisa, Largo Pontecorvo 3, 56127 Pisa, Italy}

\author{R. Donofrio\,\orcidlink{0009-0000-2789-4145}}
\altaffiliation{Present address: Universität Innsbruck, Institut für Experimentalphysik, Technikerstraße 25, 6020 Innsbruck, Austria}
\altaffiliation{Institut für Quantenoptik und Quanteninformation, Österreichische Akademie der Wissenschaften}
\affiliation{Dipartimento di Fisica ‘E. Fermi’, Universit\`a di Pisa, Largo Pontecorvo 3, 56127 Pisa, Italy}

\author{B. Bégoc\,\orcidlink{0009-0008-2906-7555}}
\altaffiliation{Present address:LTE (former SYRTE), Observatoire de Paris, Université PSL, CNRS, Sorbonne Université, LNE 
61 avenue de l’Observatoire, 75014 Paris, France}
\affiliation{Dipartimento di Fisica ‘E. Fermi’, Universit\`a di Pisa, Largo Pontecorvo 3, 56127 Pisa, Italy}

\author{O. Morsch\,\orcidlink{0000-0002-1063-5136}}
\email[Mail to: ]{oliver.morsch@cnr.it}
\affiliation{Dipartimento di Fisica ‘E. Fermi’, Universit\`a di Pisa, Largo Pontecorvo 3, 56127 Pisa, Italy}
\affiliation{CNR-INO, Via G. Moruzzi 1, 56124 Pisa, Italy}

\begin{abstract}
We demonstrate a technique for spatially resolved temperature measurement utilizing Rydberg Doppler broadening thermometry. This method employs two focused laser beams arranged perpendicularly to excite laser-cooled atoms from the ground state to a Rydberg state via a two-photon absorption process. Temperature is obtained through Doppler broadening of the spectral line. The perpendicular configuration allows for selective probing of a specific position within the atomic cloud, enabling localized temperature measurement. This technique, in principle, offers a temperature resolution on the order of \SI{}{\nano\kelvin}, attributed to the exceptionally narrow natural linewidth of the involved rubidium Rydberg transition line. 
Furthermore, the setup enables the measurement of position-velocity correlations within the cold atom ensemble. The velocity information is extracted through the Doppler shift, whereas the spatial information is inferred from the arrival time of ions detected by a channel electron multiplier detector. We use our method to measure the local temperature in a magneto-optical trap.
\end{abstract}

\maketitle

Laser cooling and trapping are well-established techniques for preparing cold samples of atoms and molecules. Such samples are essential for applications including atomic clocks \cite{atomic_clock_diddams,takamoto2005opticallatticeclock}, the creation of Bose–Einstein condensates \cite{BEC_ketterle}, and precision quantum sensing \cite{gyroscope,budker2007opticalmagnetometry}.
A large body of experimental work has investigated the properties of magneto-optical traps (MOTs)\cite{phase_space_density_foot, Radiation_force_magneto_optical_trap,MOT_properties_transient_oscillation_method, characteristics_MOT_2017,Jun_ye} . In particular, MOTs are known to exhibit rich dynamical behavior, including nonlinear dynamics \cite{nonlinear_dynamics_Labeyrie_2006,MOON20171,D_wilkowski}, phase transitions, and instabilities\cite{MOT_instability_2006,MOT_instability_2020,MOT_instability_2021}.
A key property of these ultracold ensembles is their temperature, a primary parameter that underpins coherence, their phase-space density, and the sensitivity of precision measurements.
Conventionally, temperature is measured using time-of-flight (TOF) expansion or release-and-recapture methods \cite{observation_of_atoms_below_doppler_cooling_TOF,MOT_lithium_TOF,MOT_Thulium_TOF,Viscous_confinement_TOF_Chu,Temp_release_recapture,The_Temp_of_atoms_in_MOT_CJ_FOOT,Temp_as_function_of_various_parameters}. These methods are based on the assumption of a linear position–velocity correlation of atoms and typically require a large number of atoms to achieve a good signal-to-noise ratio. 
To overcome this limitation, one can instead use spectroscopic techniques based on the Doppler effect \cite{doppler_spectroscopy}. 
The application of this technique requires a transition with a natural linewidth significantly narrower than the Doppler width. Recent advances in two-photon excitation \cite{Two_photon_absorption_selection_rule} to Rydberg states provide access to resonances with natural linewidths  well below the Doppler width (typically kHz level)\cite{Rydberg-Tagging,differentialtwo-photon2018,differentialtwo-photon2020}. This allows, in principle, for temperature measurements with nK resolution (see Supplemental Material \cite{SupplementalMaterial} for additional details). However, while previous studies have demonstrated Rydberg-based thermometry in MOTs  \cite{Rydberg-Tagging,differentialtwo-photon2018,differentialtwo-photon2020,single_photon_Doppler_free_spectroscopy}, these approaches typically averaged over the entire atomic cloud, thus discarding local spatial information essential for studying nonequilibrium features such as heat transport \cite{two_terminal_heat_transport} and Doppler heating \cite{SoleCardona2026DopplerHeating}. Recently, spatially resolved temperature measurements have also been demonstrated using Raman spectroscopy\cite{Raman_spectroscopy,Raman_spectroscopy_Sagi}. While such Raman-based methods provide a powerful and complementary approach, they generally require comparatively large excitation volumes to maintain sufficient Raman transfer efficiency, which limits their applicability for microscopic local thermometry.

Here, we demonstrate spatially selective temperature measurements of an atomic cloud of ultracold $^{87}$Rb atoms using Rydberg Doppler broadening thermometry. 
By addressing localized regions with laser beams an order of magnitude smaller than the dimensions of the atomic cloud, we can measure spatial variations in temperature across the sample.
Beyond spatial resolution, our technique offers several additional advantages. 
We demonstrate temperature measurements over a broad range, from the \SI{}{\micro\kelvin} to \SI{}{\milli\kelvin}   regime.
Moreover, by probing atoms after a period of free expansion, we measure position--velocity correlations, an observable previously studied using complementary phase-space tomography techniques\cite{position_velocity_correlation_using_Raman_spectroscopy,position_velocity_correlation_Folman}.
Our implementation provides a distinct route based on velocity-selective Rydberg excitation and field ionization detection.

\begin{figure}[htbp]
    \centering
    \includegraphics[width=\columnwidth]{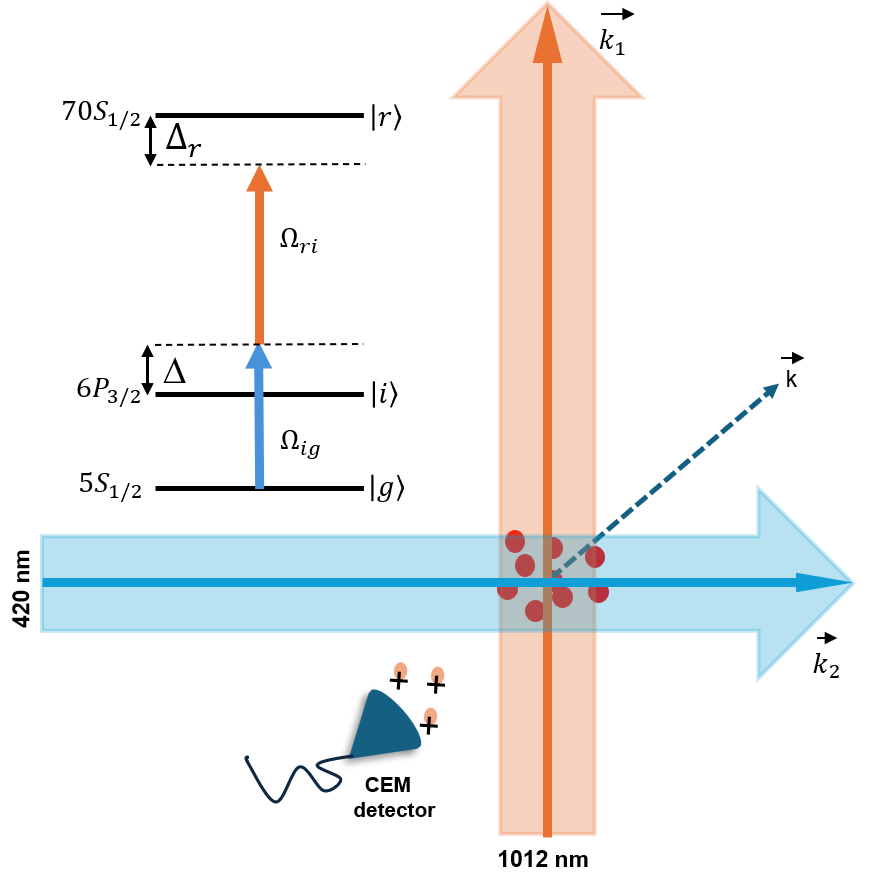} 
    \caption{\textbf{Excitation and detection scheme.} 1012 nm ($\vec{k_1}$) and 420 nm ($\vec{k_2}$) laser beams (large or small) are used in perpendicular configuration to excite the atoms into Rydberg state $\ket{r}$ from ground state $\ket{g}$ via an intermediate state $\ket{i}$.  $\Delta$, $\Delta_r$ are the detunings of 420 nm and 1012 nm laser beam respectively and $\Omega_{ig}$, $\Omega_{ri}$ are the single photon Rabi frequencies. The angle between the effective wave vector $\Vec{k}$ and  $\Vec{k_2}$ (blue laser) is $22.5^\circ$.}
    \label{fig:full experimental setup} 
\end{figure}

In our experiments, we excite laser-cooled $^{87}$Rb atoms to the $70S_{1/2}$ Rydberg state in a standard magneto-optical trap (MOT) containing $\sim10^{6}$ atoms with a $1\sigma$ radius of $\SI{160}{\micro\meter}$ (Gaussian density profile).  The excitation to the Rydberg state \cite{Rydberg_Excitations} is achieved through two-photon absorption process (Fig. \ref{fig:full experimental setup}), where atoms absorb photons from two linearly polarized laser beams at 420 nm and 1012 nm, respectively, arranged in a perpendicular configuration, to transition from $\ket{5S_{1/2}}$, F=2 to $\ket{70S_{1/2}}$ Rydberg state via the intermediate $\ket{6P_{3/2}}$, F'=3 state, detuned by +200 MHz. Depending on the measurement, we use different beam-waist configurations. For global thermometry, both excitation beams are expanded to waists of $\SI{1.6}{\milli\meter}$ (ten times the MOT radius). For position-velocity correlation measurements, we use a small $\SI{420}{\nano\meter}$ beam (waist $\SI{16}{\micro\meter}$) together with a large $\SI{1012}{\nano\meter}$ beam (waist $\SI{1.6}{\milli\meter}$), providing spatial selectivity in one dimension. For local temperature measurements, both beams are small, with waists of $\SI{16}{\micro\meter}$ and $\SI{36}{\micro\meter}$ for $\SI{420}{\nano\meter}$ and $\SI{1012}{\nano\meter}$ beams, respectively. To detect Rydberg atoms, we employ a field ionization process \cite{gallagher1994rydberg,gregoric2018improving,robicheaux1997pulsed,hollenstein2001selective,walz2004cold}. The ions produced by field ionization are accelerated toward a channel electron multiplier (CEM) using the same electric field applied for ionization. Individual ions are detected as voltage peaks on an oscilloscope connected to the CEM and counted by a peak finding procedure in LabVIEW software with an overall detection efficiency of \SI{40}{\%} \cite{Ion_detection}. The arrival times of the individual ions are also recorded and are used to infer the position of the Rydberg atoms.

The combination of using excitation beams in a perpendicular configuration and exciting atoms to Rydberg states allows us to probe the local dynamics of atoms in a small volume within the atomic cloud. Additionally, the small number of atoms in the selected region does not pose a limitation, since only a few atoms excited to Rydberg states are sufficient to achieve a good signal-to-noise ratio.
 
In a single experimental cycle, after a loading time of \SI{100}{\milli\second}, we switch off the MOT magnetic field to minimize the Zeeman broadening. After \SI{600}{\micro\second} (at this point, the quadrupole field has decayed sufficiently; see Supplemental Material \cite{SupplementalMaterial} for further details), the cooling beams are switched off and the Rydberg excitation beams are pulsed on for \SI{10}{\micro\second}. Excited atoms are then detected using field ionization. After the measurement, the cooling beams and magnetic field are switched on again to reload the MOT for the next experimental cycle. For each data point, the sequence is repeated 200 times to accumulate sufficient statistics.

\begin{figure}[t]
  \centering

  \begin{overpic}[width=0.5253\columnwidth]{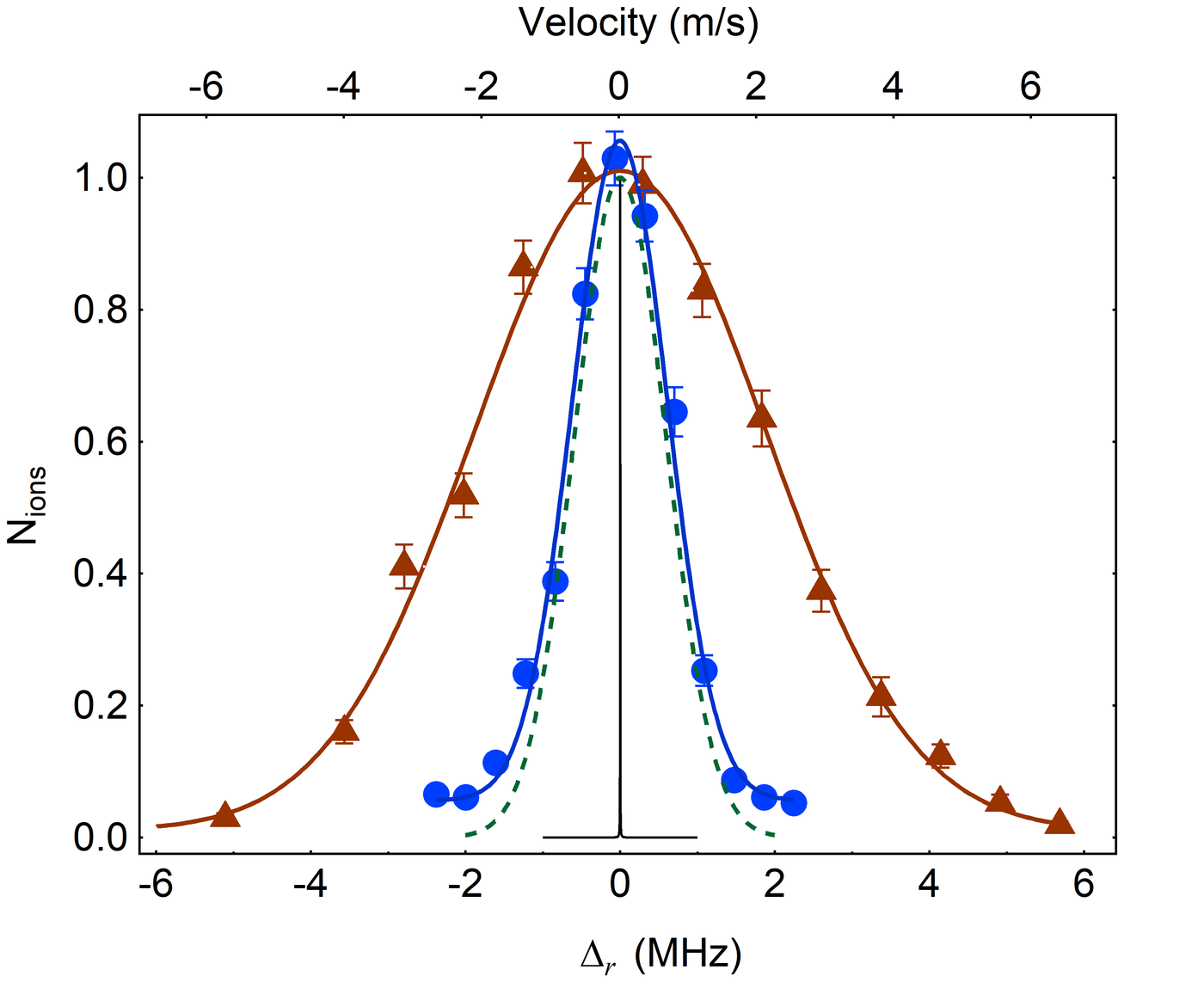}
    \put(12,66){\fontfamily{phv}\selectfont\text{(a)}}
  \end{overpic}%
  \hspace{-0.03\columnwidth}%
  \begin{overpic}[width=0.5007\columnwidth]{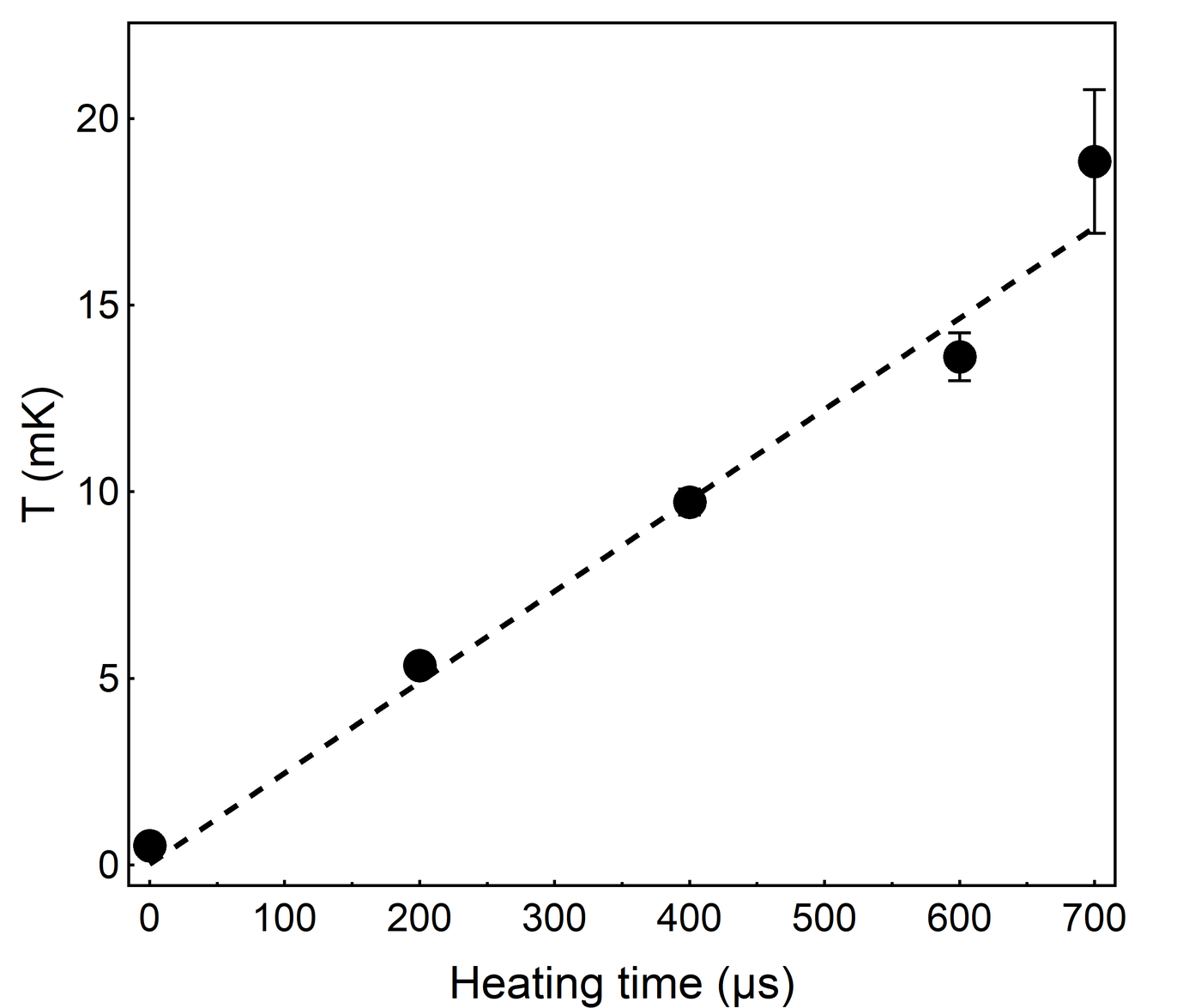}
    \put(12,75){\fontfamily{phv}\selectfont\text{(b)}}
  \end{overpic}

  \caption{\textbf{Excitation spectrum and heating rate.}
  (a) Number of Rydberg excitations ($70S_{1/2}$) as a function of excitation-laser detuning $\Delta_r$; the corresponding atomic velocities are indicated on the top axis. Triangular and circular data points correspond to \SI{4.8(2)}{\milli\kelvin} and \SI{210(50)}{\micro\kelvin}, respectively. The solid red and blue curves are Gaussian fits to the experimental data. The dashed curve shows broadening due to frequency jitter and the narrow Lorentzian function indicates the natural linewidth.
  (b) Temperature of the atomic cloud versus heating time $\Delta t$. The dashed line is a linear fit, yielding a heating rate of \SI{22.9(6)}{\kelvin\per\second}.}
  \label{fig:excitation_spectra_heating}
\end{figure}

To test our method, we first measure the global temperature of the atoms in the MOT using large excitation beams. Figure~\ref{fig:excitation_spectra_heating}(a) shows the number of Rydberg atoms as a function of the laser detuning $\Delta_r$ for two different temperatures of the atomic cloud. The experimental data are fitted with a Gaussian profile to extract the total full width at half maximum (FWHM) $\Delta f_{\mathrm{Total}}$. The measured linewidth contains contributions from Doppler broadening $\Delta f_{\mathrm D}$, and non-Doppler mechanisms, dominated by laser-frequency jitter $\Delta f_{\mathrm{L}}$ (see Supplemental Material \cite{SupplementalMaterial}). Additional contributions include power broadening, natural linewidth and Zeeman broadening in the presence of magnetic fields. In our experiments, the excitation-beam intensities are sufficiently low that power broadening can be neglected (below \(100~\mathrm{kHz}\)). The largest peak intensities used in the experiment are about \(\SI{2.5}{\watt\per\centi\meter\squared}\) for the \(420~\mathrm{nm}\) laser and \(\SI{46}{\watt\per\centi\meter\squared}\) for the \(1012~\mathrm{nm}\) laser. The natural linewidth of the Rydberg transition, \SI{0.95}{\kilo\hertz}, is also negligible. Moreover, because the magnetic field is switched off before the excitation pulse, Zeeman broadening is negligible, contributing less than \SI{90}{\kilo\hertz} as estimated from measurements of the residual magnetic-field decay. Consequently, the dominant non-Doppler broadening mechanism is the laser jitter $\Delta f_{\mathrm{L}}$, which we measured to be \SI{1.39(0.18)}{\mega\hertz} by two independent methods: (i) the standard deviation of the peak position recorded by the LabVIEW-based laser-frequency stabilization feedback loop, and (ii) a comparison of spectra taken in co- and counterpropagating beam geometries (see Supplemental Material \cite{SupplementalMaterial}). From the method (ii), we can also infer that the laser jitter can be described by a Gaussian distribution. Consequently, the total FWHM linewidth is given by the quadrature sum of the two Gaussians, the Doppler and laser broadening: 
\begin{equation}
\Delta f_{\mathrm{Total}}
=
\sqrt{\Delta f_{\mathrm D}^2 + \Delta f_{\mathrm{L}}^2}\, ,
\label{eq:total_broadening}
\end{equation}
where the Doppler contribution is
\begin{equation}
\Delta f_{\mathrm D}
=
\frac{\sqrt{\ln 2}}{\pi}\,|\vec{k}_1+\vec{k}_2|\,U.
\label{eq:doppler_broadening}
\end{equation}
Here $U$ is the most probable thermal speed,
\begin{equation}
U=\sqrt{\frac{2k_BT}{m}},
\label{eq:uth}
\end{equation}
and $|\vec{k}_1+\vec{k}_2|$ is the effective two-photon wavevector~\cite{differentialtwo-photon2018,differentialtwo-photon2020}. For two wave vectors $\vec{k}_1$ and $\vec{k}_2$ enclosing an angle $\beta$,
\begin{equation}
|\vec{k}_1+\vec{k}_2|
=
\sqrt{k_1^2+k_2^2+2k_1k_2\cos\beta}.
\label{eq:keff}
\end{equation}

Using Eq.~(\ref{eq:total_broadening}), together with the measured $\Delta f_{\mathrm{Total}}$ and the independently measured laser-broadening contribution $\Delta f_{\mathrm L}$, we obtain the Doppler width $\Delta f_{\mathrm D}$ and subsequently determine the temperature via Eq.~(\ref{eq:doppler_broadening}). 
In Fig. \ref{fig:excitation_spectra_heating}(a), two scans are shown. In the first scan, the detuning of the MOT beams is kept at standard value used for MOT operation of -2$\Gamma$ (where $\Gamma$ is natural linewidth of the transition $5S_{1/2} \rightarrow 5P_{3/2}$), resulting in a measured temperature of \SI{210(50)}{\micro\kelvin}, compatible with typical values reported for $^{87}\mathrm{Rb}$ MOTs \cite{MOT_temp1,MOT_temp2,MOT_temp3} and with the Doppler temperature $T_D$ = \SI{146}{\micro\kelvin}. The second scan was taken after switching the MOT beams to resonance for \SI{200}{\micro\second}. In this case, the atomic cloud is heated to \SI{4.8(2)}{\milli\kelvin} by photon scattering. By varying this heating time, we also demonstrate the ability of our method to measure temperature over several orders of magnitude from \SI{\sim 100}{\micro\kelvin} to \SI{20}{\milli\kelvin}, as shown in Fig. \ref{fig:excitation_spectra_heating}(b). As expected, the measured temperature increases linearly with heating time, corresponding to a heating rate of \SI{22.9(6)}{\kelvin\per\second}. A simple theoretical estimate based on scattering-rate calculations predicts a heating rate of approximately \SI{6}{\kelvin\per\second}. The difference from the experimentally observed value may be attributed to uncertainty in the exact resonance frequency of the cooling beams. In particular, a small blue detuning, on the order of \SI{200}{\kilo\hertz}, would give rise to anomalous Doppler heating \cite{doppler_heating_jun_ye, SoleCardona2026DopplerHeating}, leading to an enhanced experimental heating rate.

The accessible temperature range is bounded from below by the laser-induced linewidth. Because the linewidths can only be measured with finite precision, the Doppler contribution must exceed its propagated uncertainty to be reliably resolved. Assuming a $\SI{10}{\%}$ uncertainty in both the measured total width and the independently determined laser broadening, we estimate a minimum accessible temperature of $\sim\SI{68}{\micro\kelvin}$ for our setup. Experimentally, the lowest temperature we measure is $\SI{63(40)}{\micro\kelvin}$, obtained for an optical molasses prepared with a cooling-laser detuning of $-4\Gamma$.

\begin{figure}[htbp]
  \centering

  \begin{overpic}[width=0.5505\columnwidth]{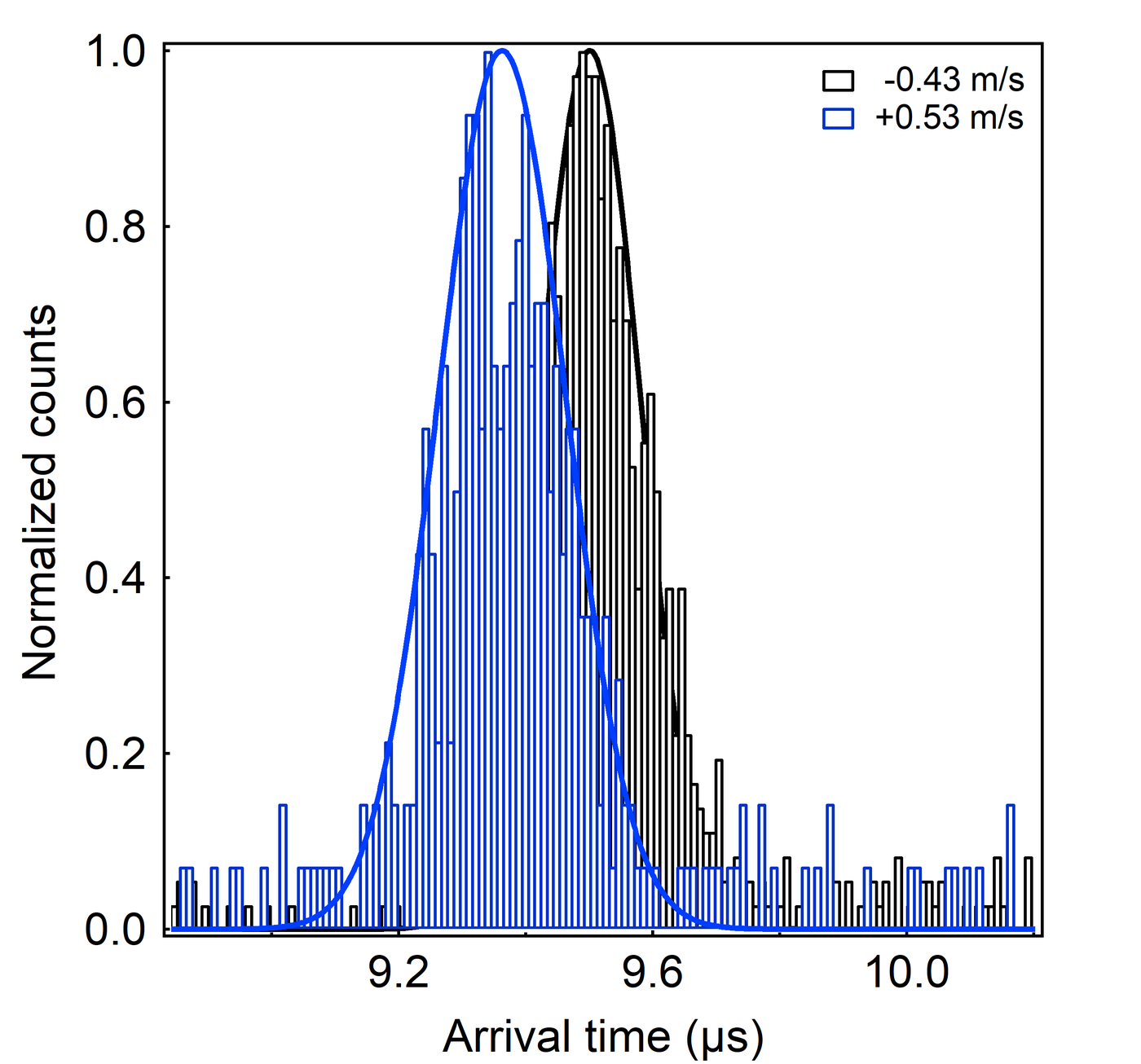}
    \put(15,83){\fontfamily{phv}\selectfont\text{(a)}}
  \end{overpic}%
  \hspace{-0.03\columnwidth}%
  \begin{overpic}[width=0.4795\columnwidth]{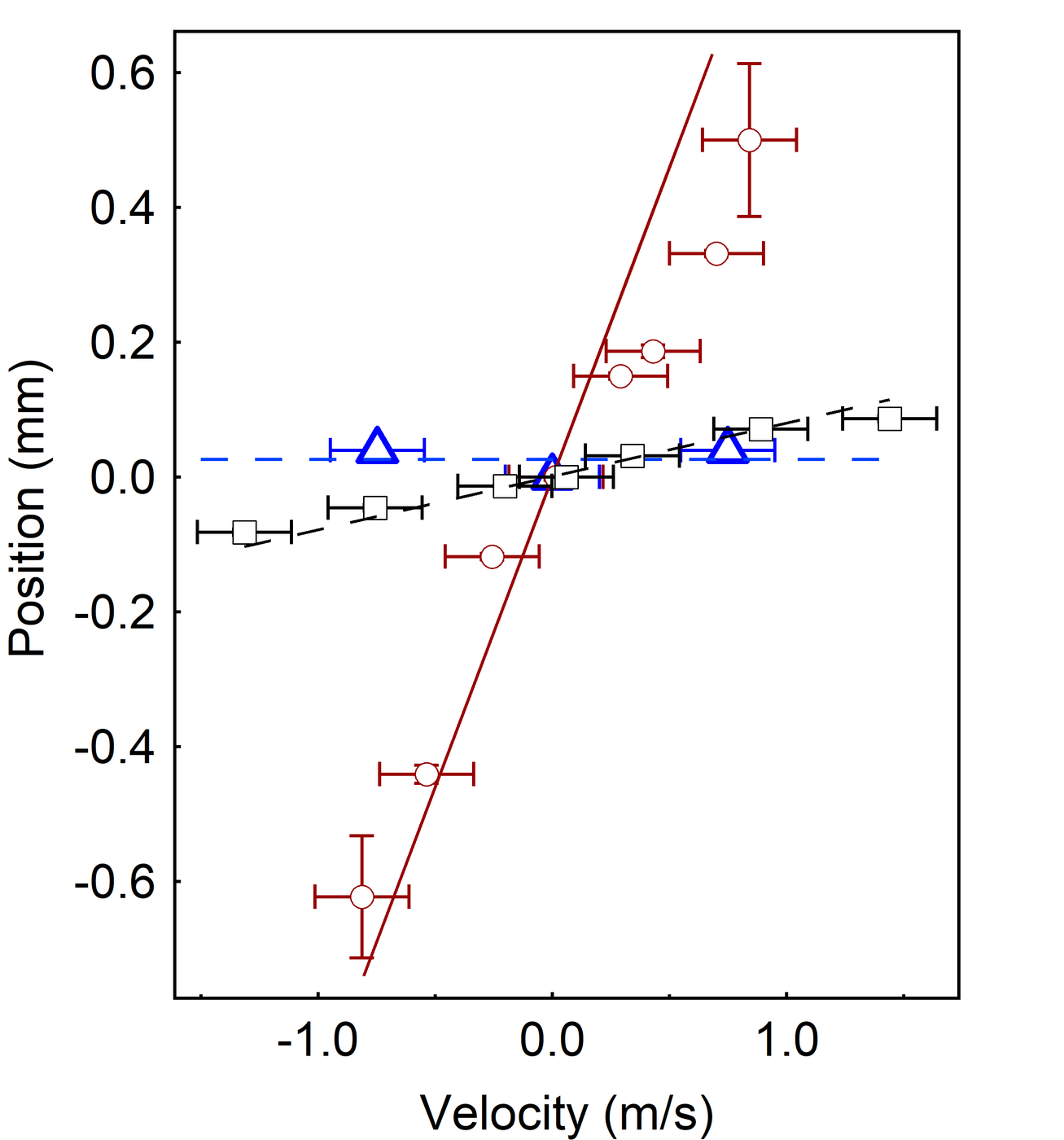}
    \put(16,90.5){\fontfamily{phv}\selectfont\text{(b)}}
  \end{overpic}

  \caption{\textbf{Position--velocity correlation measurements.}
  (a) Histogram of ion arrival times for two selected velocity classes. The measurement was performed after heating the cloud for \SI{400}{\micro\second}, followed by \SI{1}{\milli\second} of ballistic expansion. Two velocity classes, \SI{-0.43}{\meter/\second} and \SI{0.53}{\meter/\second}, corresponding to the right peak (black bars) and left peak (blue bars), originate from positions \(\SI{+0.19}{\milli\meter}\) and \(\SI{-0.44}{\milli\meter}\) relative to the cloud center, respectively.
  (b) Relative position with respect to atoms with zero velocity as a function of the velocity component projected along the \SI{420}{\nano\meter} beam, as inferred from the Doppler shift. The projected velocity was calculated by multiplying the measured velocity by $\cos(22.5^\circ)$. Triangles correspond to a sample in thermal equilibrium, while squares and circles denote the distributions measured immediately after resonant heating and after \SI{1}{\milli\second} of ballistic expansion, respectively.
  The dashed lines show linear fits to the data points, and the solid line represents the theoretically calculated position after \SI{1}{\milli\second} of expansion time.}
  \label{fig:pv_corr}
\end{figure}

Our experimental setup also allows us to infer the position of Rydberg atoms using arrival times of the ions and we can use this information to measure the position-velocity correlation of the atoms. Figure~\ref{fig:pv_corr}(a) shows the histogram of ion arrival times for two distinct velocity classes, \SI{-0.43}{\meter\per\second} and \SI{0.53}{\meter\per\second}, corresponding to $\Delta_r=\SI{-1.21}{\mega\hertz}$ and \SI{1.5}{\mega\hertz}, respectively. The data were taken after heating the cloud for \SI{400}{\micro\second}, followed by \SI{1}{\milli\second} of ballistic expansion. The peaks of the two arrival time distributions are separated in arrival time by \SI{0.14}{\micro\second}. To calibrate the mapping between ion arrival time and position, we translated the MOT by shifting the magnetic-field zero by a known distance and measure the resulting change in arrival time (see Ref.~\cite{van_der_Waals_explosion} for details). Using this calibration, we infer that the two velocity classes are separated spatially by \SI{636}{\micro\meter}. To measure the correlation, we select a specific velocity class via the Doppler shift and then measure the number of atoms in that class, as well as their arrival time. Results are shown in Fig. \ref{fig:pv_corr}(b). Triangular data points represent the measurement of the correlation without letting the cloud expand. As expected, no correlation is observed between the atomic position and velocity in the initial cloud, consistent with thermal equilibrium. After resonant heating for \SI{400}{\micro\second} (squares), a correlation begins to emerge. Letting the cloud undergo an additional \SI{1}{\milli\second} of ballistic expansion further enhances this correlation (circles) by approximately a factor of 12. The observed linear relation between position and velocity arises naturally during free expansion, as atoms with higher velocities travel farther from their initial positions. More generally, in situations where the position–velocity relation is not expected to be linear, this method can still be used to reconstruct the spatial distribution associated with different atomic velocity classes \cite{doppler_heating_jun_ye, SoleCardona2026DopplerHeating}.

\begin{figure}[htbp]
    \centering
    \includegraphics[width=\columnwidth]{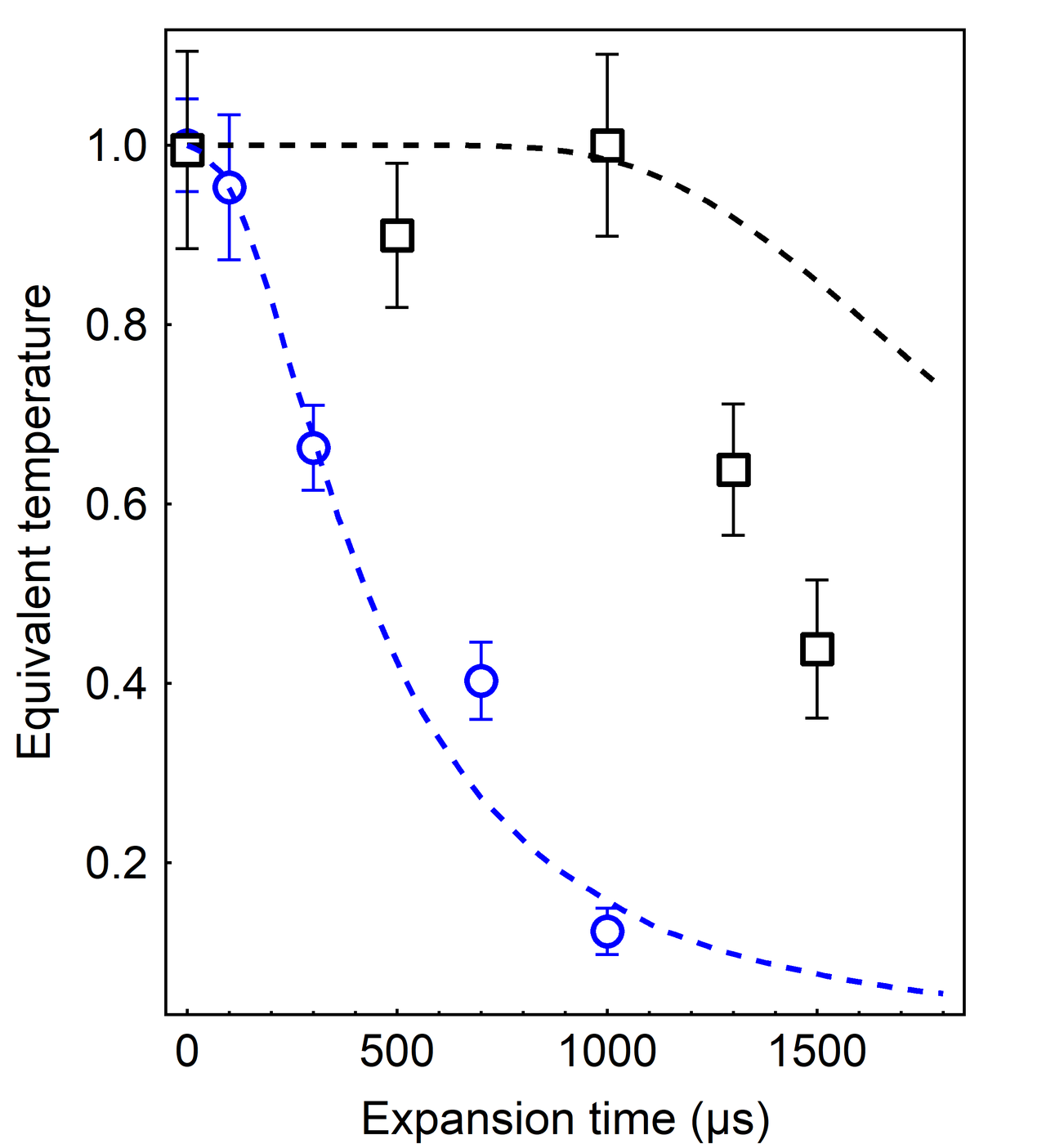}
    \caption{\textbf{Equivalent temperature as a function of expansion time.} Circular and square data points represent the equivalent temperature measurement using small and large excitation beams respectively. The temperatures are normalized to their values at zero expansion time. Dashed curves are the simulation results.}
    \label{fig:expansion cooling} 
\end{figure}

Finally, our method enables local probing of the atomic cloud using small excitation beams. To test this capability, we measured the ballistic expansion both globally and locally.
For local measurements, we define a small probing volume (an order of magnitude smaller than the atomic cloud) near the center of the atomic cloud. An atomic sample with an initial temperature of  \SI{\sim2}{\milli\kelvin}  is prepared by resonant heating, after which the MOT beams are switched off. After a variable expansion time up to \SI{1500}{\micro\second}, we measure the velocity spread in the volume. Because the expanding cloud is out of thermal equilibrium, a unique thermodynamic temperature is not well defined; we therefore report an equivalent temperature, defined from the local mean velocity.
Fig.~\ref{fig:expansion cooling} shows the equivalent temperature as a function of expansion time. For local probing (circles), the equivalent temperature decays rapidly. During ballistic expansion, only atoms with small velocity remain within the probe volume. In contrast, the global measurement with excitation beams 10 times larger than the MOT radius (squares) yields an almost constant equivalent temperature over the first millisecond, followed by a slower decay.
The dashed lines in Fig.~\ref{fig:expansion cooling} represent simulation results that show good qualitative agreement with the experimental data. In these simulations, we generate an ensemble of particles whose velocities follow a Maxwell–Boltzmann distribution and whose spatial density has a Gaussian profile. For simplicity, the excitation region is approximated by a spherical volume with a radius equal to the measured excitation beam waist, and the cloud expansion is then modeled using simple kinematic relations. This approximation could explain why there is deviation between the simulated and experimental decay of the equivalent temperature. Furthermore, uncertainty in the resonance condition could lead to Doppler heating (as discussed above)\cite{doppler_heating_jun_ye, SoleCardona2026DopplerHeating} and, hence, a larger width of the initial velocity distribution, resulting in faster decay of the measured equivalent temperature.

\begin{figure}[htbp]
    \centering
    \includegraphics[width=\columnwidth]{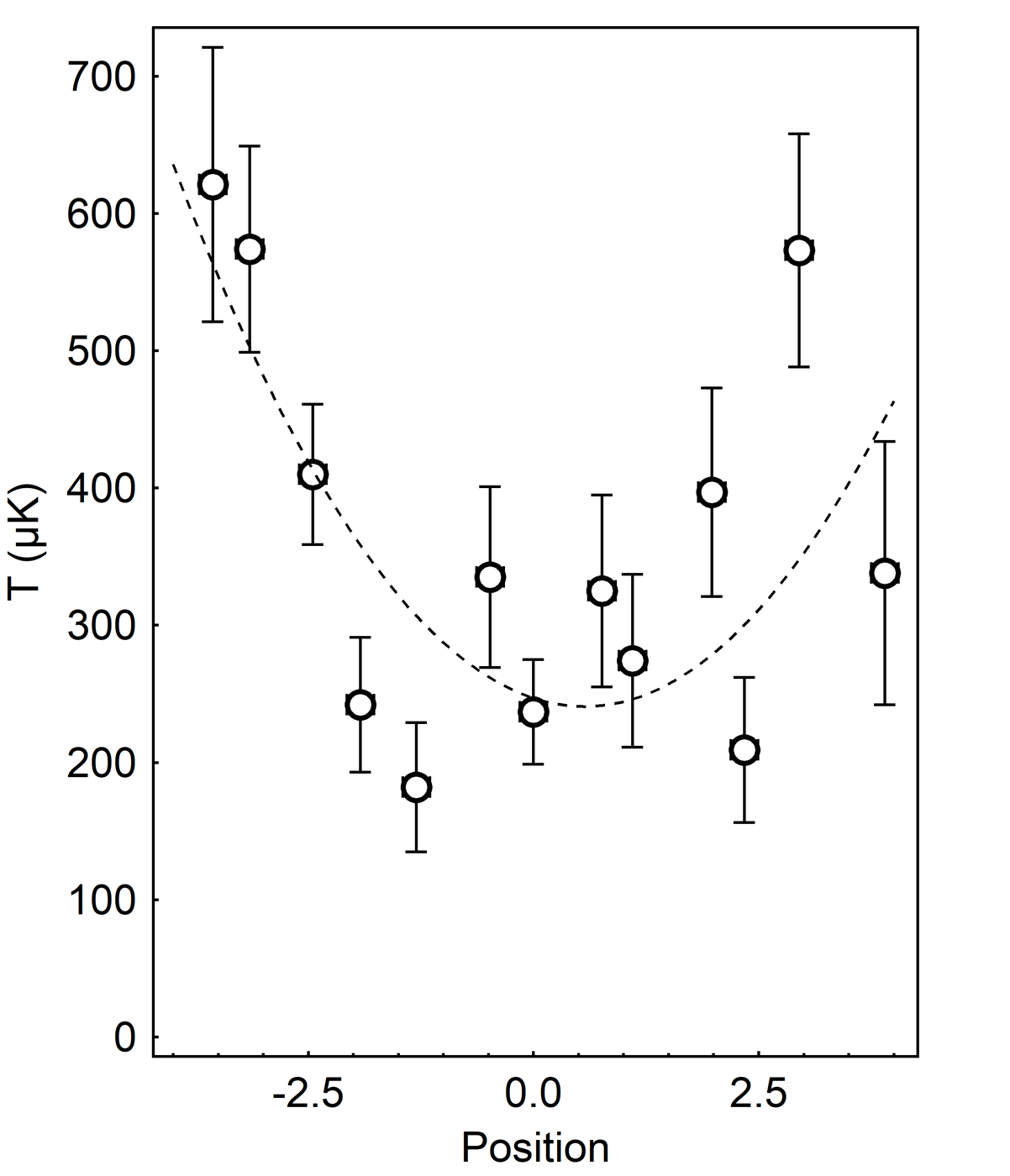} 
    \caption{\textbf{Local temperature as a function of position across the MOT.} The horizontal axis shows the position of the probed local volume in units of the MOT radius. Vertical error bars indicate the uncertainty in temperature due to error in total broadening measurements and horizontal error bars indicate the uncertainty in determining the position of the excitation laser beam. The dashed curve is included as a guide to the eye.}
    \label{fig:Temp gradient} 
\end{figure}

We then used this method to measure the temperature variation along a single dimension in the MOT using small excitation beams. A previous theoretical study \cite{arnold2000atomic} predicts that a temperature gradient must exist in the MOT for the density solution to remain physically valid. The measurement position within the MOT was selected by translating \SI{1012}{\nano\meter} excitation beam while keeping \SI{420}{\nano\meter} excitation beam fixed. The resulting beam displacement was monitored with a CCD camera positioned along the propagation direction of the translated beam. The results are shown in Fig.~\ref{fig:Temp gradient}. The MOT temperature is clearly nonuniform and exhibits a spatial gradient, with higher temperatures in the wings and lower temperatures near the center. The temperature in the wings is approximately 3 times higher than that measured in the center of the MOT. A significant temperature difference becomes apparent beyond a distance of approximately $2.5\sigma$ from the center of the cloud.

We note here that similar measurements have been performed in optical molasses using Raman-based spectroscopy \cite{Raman_spectroscopy}, which showed a similar inhomogeneous temperature distribution.

Local temperature fluctuations may arise from spatial inhomogeneities in the atomic density induced by interference fringes generated by counterpropagating cooling beams.

In conclusion, we have experimentally demonstrated that Rydberg Doppler-broadening thermometry enables precise, local measurements over a wide temperature range spanning from \SI{}{\micro\kelvin} to \SI{}{\milli\kelvin}. We use this technique to reveal position–velocity correlations through spatially resolved Rydberg Doppler thermometry, complementing previous approaches based on phase-space tomography\cite{position_velocity_correlation_using_Raman_spectroscopy, position_velocity_correlation_Folman}. To confirm the method's spatial locality, we then analyzed the ballistic expansion behavior for different excitation beam configurations. Finally, we validated the technique’s robustness by mapping the local temperature distribution across the MOT. Although the present Letter focuses on demonstrating the method, future experiments could use it to systematically investigate how local temperature gradients depend on MOT parameters, such as cooling-beam intensity, detuning, polarization, and magnetic-field gradient.

These results establish Rydberg Doppler thermometry as a powerful tool for spatially resolved thermometry in cold atom systems, with applications to nonequilibrium studies such as heat transport and Doppler-heating dynamics within the atomic cloud. Looking ahead, reducing the laser-frequency jitter that dominates the non-Doppler linewidth would lower the minimum accessible temperature and improve the resolution. In the limit of negligible laser-induced broadening, the intrinsically narrow Rydberg resonance would enable thermometry with nanokelvin resolution.

\textit{Acknowledgments---}
We gratefully acknowledge funding by the Julian Schwinger Foundation Grant No. JSF-18-1200, the H2020 ITN “MOQS” (Grant Agreement No. 955479), and the PNRR MUR Project No. PE0000023-NQSTI.

\textit{Data availability---}There are no publicly available research data or software supporting this manuscript. Requests for further information or data should be sent to the authors.

\nocite{apsrev42Control}
\bibliographystyle{apsrev4-2}
\bibliography{references}

@CONTROL{apsrev42Control,
  author = "08",
  editor = "1",
  pages  = "0",
  title  = "0",
  year   = "1"
}

@article{atomic_clock_diddams,
author = {S. A. Diddams  and Th. Udem  and J. C. Bergquist  and E. A. Curtis  and R. E. Drullinger  and L. Hollberg  and W. M. Itano  and W. D. Lee  and C. W. Oates  and K. R. Vogel  and D. J. Wineland },
title = {An Optical Clock Based on a Single Trapped {$^{199}$Hg$^{+}$} Ion},
journal = {Science},
volume = {293},
number = {5531},
pages = {825-828},
year = {2001},
}

@article{takamoto2005opticallatticeclock,
author  = {Masao Takamoto and Feng Lei Hong and Ryoichi Higashi and Hidetoshi Katori},
title   = {An Optical Lattice Clock},
journal = {Nature},
volume  = {435},
number  = {7040},
pages   = {321-324},
year    = {2005},
}

@article{BEC_ketterle,
  title = {{Bose-Einstein Condensation in a Gas of Sodium Atoms}},
  author = {Davis, K. B. and Mewes, M. -O. and Andrews, M. R. and van Druten, N. J. and Durfee, D. S. and Kurn, D. M. and Ketterle, W.},
  journal = {Phys. Rev. Lett.},
  volume = {75},
  issue = {22},
  pages = {3969--3973},
  numpages = {0},
  year = {1995},
  month = {Nov},
  publisher = {American Physical Society},
}

@article{gyroscope,
  title = {Precision Rotation Measurements with an Atom Interferometer Gyroscope},
  author = {Gustavson, T. L. and Bouyer, P. and Kasevich, M. A.},
  journal = {Phys. Rev. Lett.},
  volume = {78},
  issue = {11},
  pages = {2046--2049},
  numpages = {0},
  year = {1997},
  month = {Mar},
  publisher = {American Physical Society},
}

@article{budker2007opticalmagnetometry,
  author  = {Dmitry Budker and Michael Romalis},
  title   = {{Optical Magnetometry}},
  journal = {Nature Physics},
  volume  = {3},
  number  = {4},
  pages   = {227-234},
  year    = {2007},
}

@article{observation_of_atoms_below_doppler_cooling_TOF,
  author = {Lett, Paul D. and Watts, Richard N. and Westbrook, Christoph I. and Phillips, William D. and Gould, Phillip L. and Metcalf, Harold J.},
  title = {{Observation of atoms laser cooled below the Doppler limit}},
  journal = {Phys. Rev. Lett.},
  volume = {61},
  number = {2},
  pages = {169--172},
  year = {1988},
}

@article{MOT_lithium_TOF,
  author = {Sch{\"u}nemann, U. and Engler, H. and Zielonkowski, M. and Weidem{\"u}ller, M. and Grimm, R.},
  title = {{Magneto-optic trapping of lithium using semiconductor lasers}},
  journal = {Opt. Commun.},
  volume = {158},
  number = {1},
  pages = {263--272},
  year = {1998},
}

@article{MOT_Thulium_TOF,
  author = {Vishnyakova, G. A. and Kalganova, E. S. and Sukachev, D. D. and Fedorov, S. A. and Sokolov, A. V. and Akimov, A. V. and Kolachevsky, N. N. and Sorokin, V. N.},
  title = {Two-stage laser cooling and optical trapping of thulium atoms},
  journal = {Laser Phys.},
  volume = {24},
  number = {7},
  pages = {074018},
  year = {2014},
}

@article{Viscous_confinement_TOF_Chu,
  author = {Chu, Steven and Hollberg, L. and Bjorkholm, J. E. and Cable, Alex and Ashkin, A.},
  title = {Three-dimensional viscous confinement and cooling of atoms by resonance radiation pressure},
  journal = {Phys. Rev. Lett.},
  volume = {55},
  number = {1},
  pages = {48--51},
  year = {1985},
}

@article{Temp_release_recapture,
  author = {Russell, L. and Kumar, R. and Tiwari, V. B. and {Nic Chormaic}, S.}, 
  title = {Measurements on release--recapture of cold {$^{85}$Rb} atoms using an optical nanofibre in a magneto-optical trap},
  journal = {Opt. Commun.},
  volume = {309},
  pages = {313--317},
  year = {2013},
}

@article{The_Temp_of_atoms_in_MOT_CJ_FOOT,
  author = {Cooper, C. J. and Hillenbrand, G. and Rink, J. and Townsend, C. G. and Zetie, K. and Foot, C. J.},
  title = {The temperature of atoms in a magneto-optical trap},
  journal = {Europhys. Lett.},
  volume = {28},
  number = {6},
  pages = {397},
  year = {1994},
}

@article{Temp_as_function_of_various_parameters,
  author = {Vorozcovs, Andrejs and Weel, Matthew and Beattie, Scott and Cauchi, Saviour and Kumarakrishnan, A.},
  title = {Measurements of temperature scaling laws in an optically dense magneto-optical trap},
  journal = {J. Opt. Soc. Am. B},
  volume = {22},
  number = {5},
  pages = {943--950},
  year = {2005},
}

@article{doppler_spectroscopy,
  author = {Casa, G. and Castrillo, A. and Galzerano, G. and Wehr, R. and Merlone, A. and Di Serafino, D. and Laporta, P. and Gianfrani, L.},
  title = {{Primary gas thermometry by means of laser-absorption spectroscopy: Determination of the Boltzmann constant}},
  journal = {Phys. Rev. Lett.},
  volume = {100},
  number = {20},
  pages = {200801},
  year = {2008},
}

@article{Two_photon_absorption_selection_rule,
  author = {Bonin, Keith D. and McIlrath, Thomas J.},
  title = {Two-photon electric-dipole selection rules},
  journal = {J. Opt. Soc. Am. B},
  volume = {1},
  number = {1},
  pages = {52--55},
  year = {1984},
}

@article{Rydberg-Tagging,
  author = {Tallant, J. and Overstreet, K. R. and Schwettmann, A. and Shaffer, J. P.},
  title = {{Sub-Doppler magneto-optical trap temperatures measured using Rydberg tagging}},
  journal = {Phys. Rev. A},
  volume = {74},
  number = {2},
  pages = {023410},
  year = {2006},
}

@article{differentialtwo-photon2018,
  author = {Sautenkov, V. A. and Saakyan, S. A. and Bobrov, A. A. and Vilshanskaya, E. V. and Zelener, B. B. and Zelener, B. V.},
  title = {Differential two-photon spectroscopy for nondestructive temperature measurements of cold light atoms in a magneto-optical trap},
  journal = {J. Opt. Soc. Am. B},
  volume = {35},
  number = {7},
  pages = {1546--1551},
  year = {2018}
}

@article{differentialtwo-photon2020,
  author = {Zelener, B. B. and Vilshanskaya, E. V. and Saakyan, S. A. and Arshinova, I. D. and Bobrov, A. A. and Sautenkov, V. A. and Zelener, B. V.},
  title = {Differential two-photon spectroscopic measurements of cold atoms temperature with variable angle between probe beams},
  journal = {Laser Phys.},
  volume = {30},
  number = {2},
  pages = {025501},
  year = {2020}
}

@article{doppler_heating_jun_ye,
  author = {Loftus, Thomas H. and Ido, Tetsuya and Ludlow, Andrew D. and Boyd, Martin M. and Ye, Jun},
  title = {Narrow Line Cooling: Finite Photon Recoil Dynamics},
  journal = {Phys. Rev. Lett.},
  volume = {93},
  issue = {7},
  numpages = {4},
  pages = {073003},
  year = {2004},
  publisher = {American Physical Society},
}

@article{single_photon_Doppler_free_spectroscopy,
  author = {Clausen, Gloria and Scheidegger, Simon and Agner, Josef A. and Schmutz, Hansj{\"u}rg and Merkt, Fr{\'e}d{\'e}ric},
  title = {Imaging-assisted single-photon {D}oppler-free laser spectroscopy and the ionization energy of metastable triplet helium},
  journal = {Phys. Rev. Lett.},
  volume = {131},
  number = {10},
  pages = {103001},
  year = {2023},
}

@article{two_terminal_heat_transport,
  author = {Krinner, Sebastian and Esslinger, Tilman and Brantut, Jean-Philippe},
  title = {Two-terminal transport measurements with cold atoms},
  journal = {J. Phys.: Condens. Matter},
  volume = {29},
  number = {34},
  pages = {343003},
  year = {2017},
}

@article{Rydberg_Excitations,
  author = {Viteau, M. and Radogostowicz, J. and Bason, M. G. and Malossi, N. and Ciampini, D. and Morsch, O. and Arimondo, E.},
  title = {Rydberg spectroscopy of a {Rb} {MOT} in the presence of applied or ion-created electric fields},
  journal = {Opt. Express},
  volume = {19},
  number = {7},
  pages = {6007--6019},
  year = {2011},
}

@book{gallagher1994rydberg,
  author = {Gallagher, Thomas F.},
  title = {{Rydberg Atoms}},
  publisher = {Cambridge University Press},
  address = {Cambridge},
  year = {1994}
}

@article{gregoric2018improving,
  author = {Gregoric, Vincent C. and Bennett, Jason J. and Gualtieri, Bianca R. and Kannad, Ankitha and Liu, Zhimin Cheryl and Rowley, Zoe A. and Carroll, Thomas J. and Noel, Michael W.},
  title = {Improving the state selectivity of field ionization with quantum control},
  journal = {Phys. Rev. A},
  volume = {98},
  number = {6},
  pages = {063404},
  year = {2018},
}

@article{robicheaux1997pulsed,
  author = {Robicheaux, F.},
  title = {{Pulsed field ionization of Rydberg atoms}},
  journal = {Phys. Rev. A},
  volume = {56},
  number = {5},
  pages = {R3358--R3361},
  year = {1997},
}

@article{hollenstein2001selective,
  author = {Hollenstein, U. and Seiler, R. and Schmutz, H. and Andrist, M. and Merkt, F.},
  title = {{Selective field ionization of high Rydberg states: Application to zero-kinetic-energy photoelectron spectroscopy}},
  journal = {J. Chem. Phys.},
  volume = {115},
  number = {12},
  pages = {5461--5469},
  year = {2001},
}

@article{walz2004cold,
  author = {Walz-Flannigan, A. and Guest, J. R. and Choi, J.-H. and Raithel, G.},
  title = {Cold Rydberg-gas dynamics},
  journal = {Phys. Rev. A},
  volume = {69},
  number = {6},
  pages = {063405},
  year = {2004},
}

@article{Ion_detection,
  author = {Viteau, M. and Radogostowicz, J. and Chotia, A. and Bason, M. G. and Malossi, N. and Fuso, F. and Ciampini, D. and Morsch, O. and Ryabtsev, I. I. and Arimondo, E.},
  title = {Ion detection in the photoionization of a {Rb} {B}ose--{E}instein condensate},
  journal = {J. Phys. B: At. Mol. Opt. Phys.},
  volume = {43},
  number = {15},
  pages = {155301},
  year = {2010},
}

@article{van_der_Waals_explosion,
  author = {Faoro, R. and Simonelli, C. and Archimi, M. and Masella, G. and Valado, M. M. and Arimondo, E. and Mannella, R. and Ciampini, D. and Morsch, O.},
  title = {van der {W}aals explosion of cold {R}ydberg clusters},
  journal = {Phys. Rev. A},
  volume = {93},
  number = {3},
  pages = {030701},
  year = {2016},
}

@article{arnold2000atomic,
  author = {Arnold, A. S. and Manson, P. J.},
  title = {Atomic density and temperature distributions in magneto-optical traps},
  journal = {J. Opt. Soc. Am. B},
  volume = {17},
  number = {4},
  pages = {497--506},
  year = {2000},
}

@article{MOT_temp1,
author = {C. D. Wallace and T. P. Dinneen and K. Y. N. Tan and A. Kumarakrishnan and P. L. Gould and J. Javanainen},
journal = {J. Opt. Soc. Am. B},
number = {5},
pages = {703--711},
publisher = {Optica Publishing Group},
title = {Measurements of temperature and spring constant in a magneto-optical trap},
volume = {11},
month = {May},
year = {1994},
}

@article{MOT_temp2,
author = {Marcin Witkowski and Bart{\l}omiej Nag\'{o}rny and Rodolfo Munoz-Rodriguez and Roman Ciury{\l}o and Piotr Szymon \.{Z}uchowski and S{\l}awomir Bilicki and Marcin Piotrowski and Piotr Morzy\'{n}ski and Micha{\l} Zawada},
journal = {Opt. Express},
number = {4},
pages = {3165--3179},
publisher = {Optica Publishing Group},
title = {Dual {Hg}-{Rb} magneto-optical trap},
volume = {25},
month = {Feb},
year = {2017},
}

@article{MOT_temp3,
  author  = {Isichenko, Andrei and Chauhan, Nitesh and Bose, Debapam and Wang, Jiawei and Kunz, Paul D. and Blumenthal, Daniel J.},
  title   = {Photonic integrated beam delivery for a rubidium 3D magneto-optical trap},
  journal = {Nat. Commun.},
  volume  = {14},
  pages   = {3080},
  year    = {2023},
}

@unpublished{SoleCardona2026DopplerHeating,
  author = {\`Elia Sol\'e Cardona and K. N. Trivedi and M. Carminati and O. Morsch},
  title  = {Doppler heating in cold atoms},
  note   = {Manuscript in preparation},
  year   = {2026}
}

@misc{SupplementalMaterial,
  note = {See Supplemental Material at http://link.aps.org/supplemental/10.1103/kbfx-l9zk for details on the measurement of the decay time of the induced magnetic field, the extra FWHM broadening due to laser-frequency jitter, and the calculation of the minimum resolvable temperature.}
}

@article{Raman_spectroscopy,
author = {Kanxing Weng and Cheng Chu and Junjie Tang and Mengxin Zheng and Peishuang Ding and Qi Wang and Jinglong Bian and Wenfeng He and Zecheng Zhang and Bing Cheng and Bin Wu and Xiaolong Wang and Qiang Lin},
journal = {Opt. Express},
number = {19},
pages = {39510--39521},
publisher = {Optica Publishing Group},
title = {Spatial mapping of cold atom clouds using velocity-selective Raman pulses in differential atom interferometers},
volume = {33},
month = {Sep},
year = {2025},
}

@article{position_velocity_correlation_using_Raman_spectroscopy,
  title = {Observing Power-Law Dynamics of Position-Velocity Correlation in Anomalous Diffusion},
  author = {Afek, Gadi and Coslovsky, Jonathan and Courvoisier, Arnaud and Livneh, Oz and Davidson, Nir},
  journal = {Phys. Rev. Lett.},
  volume = {119},
  issue = {6},
  pages = {060602},
  numpages = {5},
  year = {2017},
  month = {Aug},
  publisher = {American Physical Society},
}

@article{position_velocity_correlation_Folman,
  title = {Phase space tomography of cold-atom dynamics in a weakly corrugated potential},
  author = {Zhou, Shuyu and Chab\'e, Julien and Salem, Ran and David, Tal and Groswasser, David and Keil, Mark and Japha, Yonathan and Folman, Ron},
  journal = {Phys. Rev. A},
  volume = {90},
  issue = {3},
  pages = {033620},
  numpages = {8},
  year = {2014},
  month = {Sep},
  publisher = {American Physical Society},
}

@article{Raman_spectroscopy_Sagi,
  title = {In situ momentum-distribution measurement of a quantum degenerate Fermi gas using Raman spectroscopy},
  author = {Shkedrov, Constantine and Ness, Gal and Florshaim, Yanay and Sagi, Yoav},
  journal = {Phys. Rev. A},
  volume = {101},
  issue = {1},
  pages = {013609},
  numpages = {5},
  year = {2020},
  month = {Jan},
  publisher = {American Physical Society},
}

@article{MOT_instability_2021,
  title = {Phase diagram of spatiotemporal instabilities in a large magneto-optical trap},
  author = {Gaudesius, M. and Zhang, Y.-C. and Pohl, T. and Kaiser, R. and Labeyrie, G.},
  journal = {Phys. Rev. A},
  volume = {103},
  issue = {4},
  pages = {L041101},
  numpages = {5},
  year = {2021},
  month = {Apr},
  publisher = {American Physical Society},
}

@article{MOT_instability_2020,
  title = {Instability threshold in a large balanced magneto-optical trap},
  author = {Gaudesius, M. and Kaiser, R. and Labeyrie, G. and Zhang, Y.-C. and Pohl, T.},
  journal = {Phys. Rev. A},
  volume = {101},
  issue = {5},
  pages = {053626},
  numpages = {7},
  year = {2020},
  month = {May},
  publisher = {American Physical Society},
}

@article{MOT_instability_2006,
  title = {Self-Sustained Oscillations in a Large Magneto-Optical Trap},
  author = {Labeyrie, G. and Michaud, F. and Kaiser, R.},
  journal = {Phys. Rev. Lett.},
  volume = {96},
  issue = {2},
  pages = {023003},
  numpages = {4},
  year = {2006},
  month = {Jan},
  publisher = {American Physical Society},
}

@article{nonlinear_dynamics_Labeyrie_2006,
  title = {Self-driven nonlinear dynamics in magneto-optical traps},
  author = {Pohl, T. and Labeyrie, G. and Kaiser, R.},
  journal = {Phys. Rev. A},
  volume = {74},
  issue = {2},
  pages = {023409},
  numpages = {4},
  year = {2006},
  month = {Aug},
  publisher = {American Physical Society},
}

@article{Jun_ye,
  title = {Dynamics in a two-level atom magneto-optical trap},
  author = {Xu, Xinye and Loftus, Thomas H. and Smith, Matthew J. and Hall, John L. and Gallagher, Alan and Ye, Jun},
  journal = {Phys. Rev. A},
  volume = {66},
  issue = {1},
  pages = {011401(R)},
  numpages = {4},
  year = {2002},
  month = {Jul},
  publisher = {American Physical Society},
}

@article{MOON20171,
title = {Nonlinear, nonequilibrium and collective dynamics in a periodically modulated cold atom system},
journal = {Physics Reports},
volume = {698},
pages = {1-30},
year = {2017},
issn = {0370-1573},
author = {Geol Moon and Myoung Sun Heo and Yonghee Kim and Heung Ryoul Noh and Wonho Jhe},
}

@article{D_wilkowski,
  title = {Linear and nonlinear magneto-optical rotation on the narrow strontium intercombination line},
  author = {Pandey, K. and Kwong, C. C. and Pramod, M. S. and Wilkowski, D.},
  journal = {Phys. Rev. A},
  volume = {93},
  issue = {5},
  pages = {053428},
  numpages = {6},
  year = {2016},
  month = {May},
  publisher = {American Physical Society},
}

@article{phase_space_density_foot,
  title = {Phase-space density in the magneto-optical trap},
  author = {Townsend, C. G. and Edwards, N. H. and Cooper, C. J. and Zetie, K. P. and Foot, C. J. and Steane, A. M. and Szriftgiser, P. and Perrin, H. and Dalibard, J.},
  journal = {Phys. Rev. A},
  volume = {52},
  issue = {2},
  pages = {1423--1440},
  numpages = {0},
  year = {1995},
  month = {Aug},
  publisher = {American Physical Society},
}

@article{Radiation_force_magneto_optical_trap,
author = {A. M. Steane and M. Chowdhury and C. J. Foot},
journal = {J. Opt. Soc. Am. B},
number = {12},
pages = {2142--2158},
publisher = {Optica Publishing Group},
title = {Radiation force in the magneto-optical trap},
volume = {9},
month = {Dec},
year = {1992},
}

@article{MOT_properties_transient_oscillation_method,
  title = {Measurement of the trap properties of a magneto-optical trap by a transient oscillation method},
  author = {Kim, Kihwan and Lee, Ki-Hwan and Heo, Myungseon and Noh, Heung-Ryoul and Jhe, Wonho},
  journal = {Phys. Rev. A},
  volume = {71},
  issue = {5},
  pages = {053406},
  numpages = {5},
  year = {2005},
  month = {May},
  publisher = {American Physical Society},
}

@article{characteristics_MOT_2017,
year = {2017},
month = {nov},
publisher = {IOP Publishing},
volume = {19},
number = {11},
pages = {113035},
author = {Williams, H J and Truppe, S and Hambach, M and Caldwell, L and Fitch, N J and Hinds, E A and Sauer, B E and Tarbutt, M R},
title = {Characteristics of a magneto-optical trap of molecules},
journal = {New Journal of Physics},
}

\clearpage
\onecolumngrid

\begin{center}
{\large \textbf{Supplemental Material to ``Spatially Resolved Temperature Measurement Using Rydberg Doppler Broadening Thermometry''}}\\[0.5em]
{\normalsize K. N. Trivedi, M. Carminati, Èlia Solé Cardona, T. Bonaccorsi, R. Donofrio, B. Bégoc, and O. Morsch}
\end{center}

\renewcommand{\thefigure}{S\arabic{figure}}
\renewcommand{\thetable}{S\arabic{table}}
\renewcommand{\theequation}{S\arabic{equation}}
\setcounter{figure}{0}
\setcounter{table}{0}
\setcounter{equation}{0}
\setcounter{section}{0}

\renewcommand{\thesection}{\Roman{section}}
\renewcommand{\thesubsection}{\Alph{subsection}}
\renewcommand{\thesubsubsection}{\arabic{subsubsection}}
\setcounter{secnumdepth}{3}


\twocolumngrid

\section{Measurement of the decay time of the induced magnetic field}
\label{app:magnetic field}

All Rydberg excitation spectra were recorded \SI{600}{\micro\second} after switching off the MOT magnetic field. In the presence of the field, the transition linewidth is additionally broadened as a result of the splitting of Zeeman sublevels, which are all populated in the MOT. However, a rapid switch-off of the MOT field creates an induced magnetic fields that decay exponentially over time.

To characterize this decay, we measured the excitation spectrum at different waiting times after the MOT magnetic field was switched off, using small excitation beams probing a small volume located \SI{428}{\micro\meter} from the cloud center.

\begin{figure}[t]
  \centering

  \begin{overpic}[width=\columnwidth]{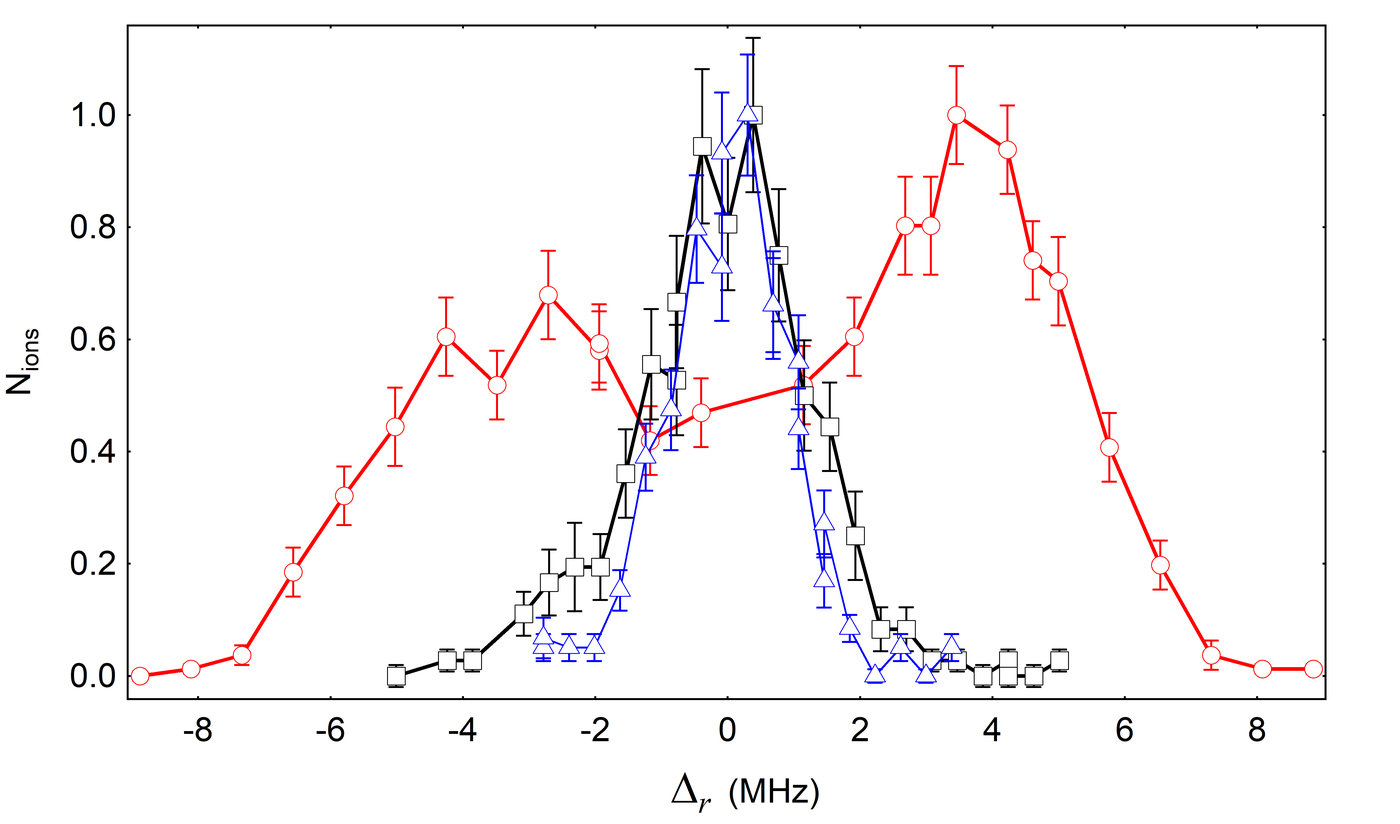}
    \put(10,53){{\fontfamily{phv}\selectfont\text{(a)}}}
  \end{overpic}

  \vspace{0.5em}

  \begin{overpic}[width=\columnwidth]{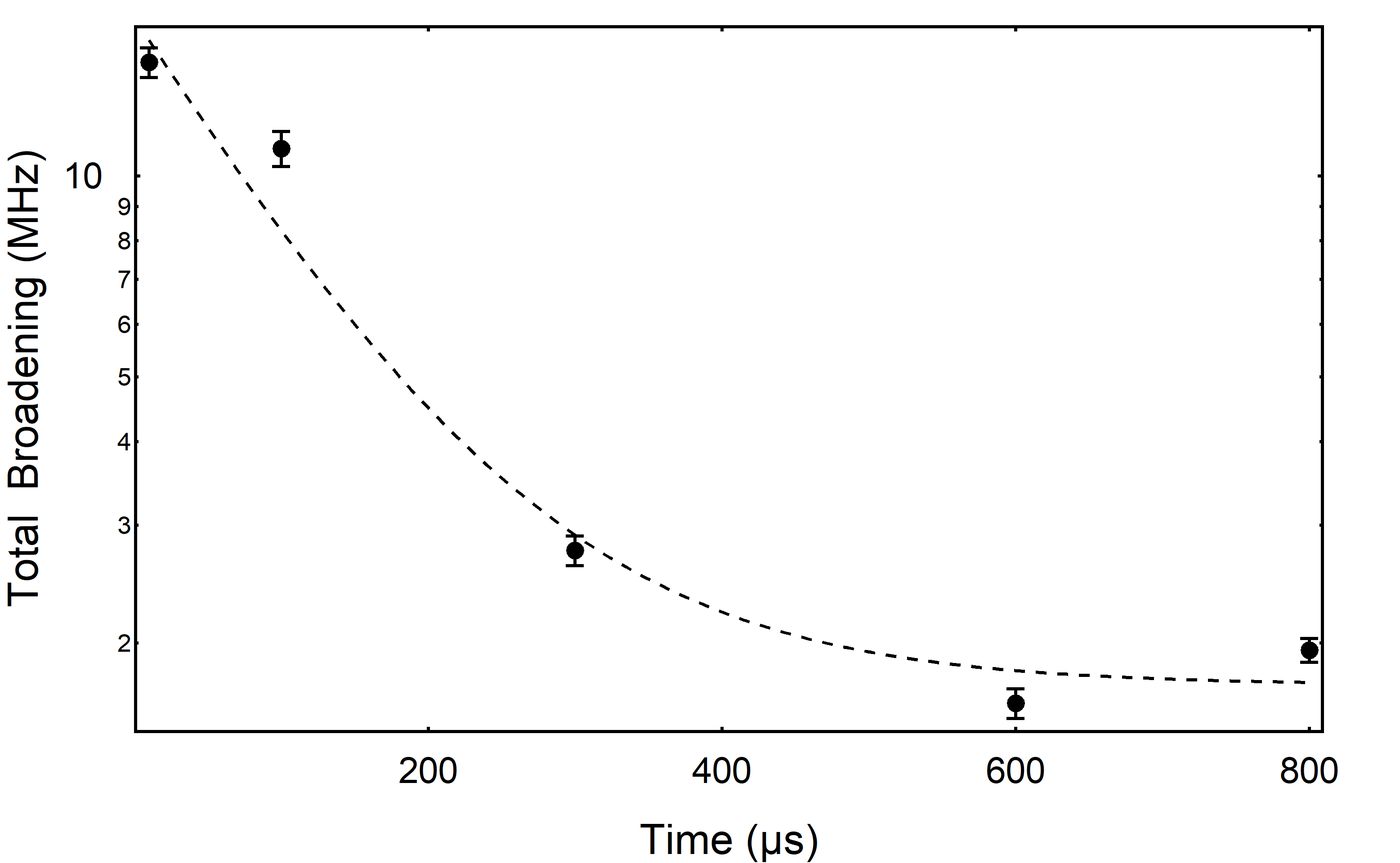}
    \put(88.5,56){{\fontfamily{phv}\selectfont\text{(b)}}}
  \end{overpic}

  \caption{
  (a) Excitation spectra for different waiting times after switching off the magnetic field. The total broadening is obtained by fitting each spectrum with a Gaussian function. Circular, square, and triangular data points represent waiting times of \SI{100}{\micro\second}, \SI{300}{\micro\second}, and \SI{800}{\micro\second}, respectively.
  (b) Total broadening as a function of waiting time after switching off the magnetic field. Vertical axis is shown in logarithmic scale. The dashed line shows an exponential fit with an offset to the extracted data points. From this fit, we obtain a decay time of the induced magnetic field of \SI{116(7)}{\micro\second}.}
  \label{fig:magnetic field decay}
\end{figure}

For waiting times below \SI{300}{\micro\second}, the excitation spectra display multiple peaks arising from Zeeman sublevel splitting [Fig.~\ref{fig:magnetic field decay}(a)]. However, to extract the FWHM in a uniform way for all waiting times, we fitted each spectrum with a single Gaussian function.
The extracted FWHM values are plotted as a function of time and fitted with an exponential decay with an offset as shown in Fig.~\ref{fig:magnetic field decay}(b). The fit yields a decay time of the induced magnetic field of approximately \SI{116(7)}{\micro\second}.
After \SI{600}{\micro\second}, the initial broadening of \SI{14.79(75)}{\mega\hertz} decreases to below \SI{1.82(7)}{\mega\hertz} as obtained from the fit function. Given the fitted offset of \SI{1.73(6)}{\mega\hertz}, the residual broadening due to the induced magnetic field at this time is around \SI{90}{\kilo\hertz}.
The fitted offset itself arises from other contributions to the linewidth, primarily Doppler broadening and laser-frequency jitter.

\section{Laser broadening measurement}
\label{app:Laser broadening measurement}

As noted in the main text, the total broadening of the Rydberg transition is
mainly determined by two contributions: the Doppler broadening,
$\Delta f_{\text{D}}$, and broadening due to frequency
jitter $\Delta f_{\text{L}}$ of the excitation lasers.

To stabilize the laser frequencies, we employ a combination of a Fabry–Pérot cavity and a LabVIEW-based control system. The excitation lasers are locked with respect to a reference laser using the cavity. The transmitted signal from the Fabry–Pérot cavity is detected by a photodiode, converting it into an electrical signal, which is then digitized using an an analog-digital converter (ADC) on a National Instruments data acquisition card.

The residual frequency jitter in the lasers arises mainly from the finite sampling and conversion rate of the ADC, which limits the feedback bandwidth of the locking system.

Although diode lasers typically exhibit a relatively narrow intrinsic
Lorentzian linewidth at the kHz level, the broadening induced by
frequency jitter is much larger and therefore
dominates over the intrinsic linewidth. Because this laser-induced
broadening is significant, we determine it experimentally using a differential measurement inspired by
Refs.~\cite{differentialtwo-photon2020,differentialtwo-photon2018}.

\begin{figure}[htbp]
    \centering
    \includegraphics[width=0.37\textwidth]{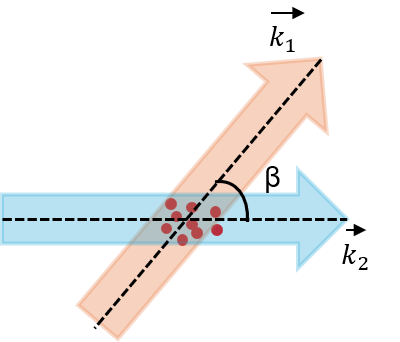}
    \caption{Excitation beam configuration used for the
    laser broadening measurement. $\beta$ is the angle between the two
    excitation laser beams. $\vec{k}_1$ and $\vec{k}_2$ are the wave
    vectors of the 1012~nm and 420~nm laser beams, respectively.}
    \label{fig:co and counter}
\end{figure}

To measure the broadening caused by frequency jitter, we use the
following protocol. The total broadening of the Rydberg transition is
measured for two different geometries of the excitation laser beams
(Fig.~\ref{fig:co and counter}). First, the measurement is performed in
a co-propagating configuration ($\beta = 0^\circ$), and then in a
counter-propagating configuration ($\beta = 180^\circ$). Since the
Doppler broadening generally depends on the angle $\beta$ according to
\begin{equation}
    |k_1 + k_2|
    =
    \sqrt{k_1^2 + k_2^2 + 2 k_1 k_2 \cos \beta},
    \label{eq:angle dependence}
\end{equation}

the two configurations yield different total broadenings:
$\Delta f_{\text{Total,co}} = \SI{2.39(2)}{\mega\hertz}$ and
$\Delta f_{\text{Total,counter}} = \SI{1.63(2)}{\mega\hertz}$.

From Eq.~\ref{eq:angle dependence}, it follows that in the
counter-propagating configuration the Doppler width is small ($\approx 390\,\mathrm{kHz}$) for
the typical temperatures of our MOT. In this case, the total
broadening is therefore dominated by the contribution arising from the
frequency jitter of the excitation lasers. Since the measured spectrum
is well described by a Gaussian fit, we assume that the broadening due
to frequency jitter is also Gaussian. Under this assumption, the total
broadening can be written as the quadrature sum of the Doppler
broadening, $\Delta f_{\text{D}}$, and the laser-frequency-jitter
broadening, $\Delta f_{\text{L}}$:
\begin{equation}
    \Delta f_{\text{Total}}
    =
    \sqrt{\Delta f_{\text{D}}^2 + \Delta f_{\text{L}}^2},
    \label{eq:total broadening}
\end{equation}

Measuring the total broadening in these two beam configurations allows
us to extract both the Doppler broadening and the frequency-jitter
broadening, $\Delta f_{\text{L}}$, by solving the following two
equations:
\begin{equation}
     \Delta f_{\text{Total,co}}^2
     =
     \Delta f_{\text{L}}^2
     +
     \frac{8 k_B T \ln 2}{m c^2}(k_1 + k_2)^2
     \label{eq:copropagating}
\end{equation}

\begin{equation}
    \Delta f_{\text{Total,counter}}^2
    =
    \Delta f_{\text{L}}^2
    +
    \frac{8 k_B T \ln 2}{m c^2}(k_1 - k_2)^2
    \label{eq:counterpropagating}
\end{equation}

Subtracting the two equations yields
\begin{equation}
    T
    =
    \frac{m c^2}{8 k_B \ln 2}
    \frac{
    \Delta f_{\text{Total,co}}^2
    -
    \Delta f_{\text{Total,counter}}^2
    }{4 k_1 k_2}.
\end{equation}

By substituting the extracted temperature $T$ into either
Eq.~\ref{eq:copropagating} or Eq.~\ref{eq:counterpropagating}, we obtain
the broadening due to frequency jitter:
\begin{equation}
   \Delta f_{\text{L}}^2
   =
   \Delta f_{\text{Total,co}}^2
   -
   \frac{
   \Delta f_{\text{Total,co}}^2
   -
   \Delta f_{\text{Total,counter}}^2
   }{4 k_1 k_2}(k_1 + k_2)^2.
\end{equation}

Using this method, we obtain a frequency-jitter broadening of
\SI{1.39(18)}{\mega\hertz}. We compare this result with an independent
estimate of the laser frequency jitter obtained from the standard
deviation of the peak position recorded by the LabVIEW-based laser
frequency-stabilization feedback loop which gives \SI{1.37(12)}{\mega\hertz}. The two values are in good agreement.

\section{Minimum resolvable temperature}
\label{app:Minimum resolvable temperature}
The Doppler width can be written as
\begin{equation}
\Delta f_{\mathrm D}=\sqrt{\Delta f_{\mathrm{Total}}^2-\Delta f_{\mathrm{L}}^2}.
\label{eq:doppler_extract}
\end{equation}
Propagating uncertainties in both $\Delta f_{\mathrm{Total}}$ and $\Delta f_{\mathrm{L}}$ gives
\begin{align}
\sigma^2(\Delta f_{\mathrm D})
&\simeq
\left(\frac{\partial \Delta f_{\mathrm D}}{\partial \Delta f_{\mathrm{Total}}}\right)^2
\sigma^2_{\mathrm{Total}}
+
\left(\frac{\partial \Delta f_{\mathrm D}}{\partial \Delta f_{\mathrm{L}}}\right)^2
\sigma^2_{\mathrm{L}}
\nonumber\\
&=
\left(\frac{\Delta f_{\mathrm{Total}}}{\Delta f_{\mathrm D}}\right)^2 \sigma^2_{\mathrm{Total}}
+
\left(\frac{\Delta f_{\mathrm{L}}}{\Delta f_{\mathrm D}}\right)^2 \sigma^2_{\mathrm{L}},
\label{eq:error_prop_both}
\end{align}
where we used
$\partial \Delta f_{\mathrm D}/\partial \Delta f_{\mathrm{Total}}=\Delta f_{\mathrm{Total}}/\Delta f_{\mathrm D}$
and
$\partial \Delta f_{\mathrm D}/\partial \Delta f_{\mathrm{L}}=-\Delta f_{\mathrm{L}}/\Delta f_{\mathrm D}$.

Writing the uncertainties as fractional errors,
$\sigma_{\mathrm{Total}}=\epsilon_{\mathrm{T}}\Delta f_{\mathrm{Total}}$ and
$\sigma_{\mathrm{L}}=\epsilon_{\mathrm{L}}\Delta f_{\mathrm{L}}$, we obtain
\begin{equation}
\sigma(\Delta f_{\mathrm D})
=
\frac{1}{\Delta f_{\mathrm D}}
\sqrt{\epsilon_{\mathrm{T}}^2\,\Delta f_{\mathrm{Total}}^4
+
\epsilon_{\mathrm{L}}^2\,\Delta f_{\mathrm{L}}^4 }.
\label{eq:sigma_doppler}
\end{equation}

A convenient ``minimum resolvable'' criterion is that the Doppler width exceeds its 1$\sigma$ uncertainty,
$\Delta f_{\mathrm D} \gtrsim \,\sigma(\Delta f_{\mathrm D})$, which implies
\begin{equation}
\Delta f_{\mathrm D}^4 \gtrsim
\!\left[
\epsilon_{\mathrm{T}}^2\,\Delta f_{\mathrm{Total}}^4
+
\epsilon_{\mathrm{L}}^2\,\Delta f_{\mathrm{L}}^4
\right].
\label{eq:resolvability_condition}
\end{equation}
In the low-temperature limit where $\Delta f_{\mathrm D}\ll \Delta f_{\mathrm{L}}$ (so that
$\Delta f_{\mathrm{Total}}\simeq \Delta f_{\mathrm{L}}$), this simplifies to
\begin{equation}
\Delta f_{\mathrm D,\min}
\simeq
\Delta f_{\mathrm{L}}\;
\left(\epsilon_{\mathrm{T}}^2+\epsilon_{\mathrm{L}}^2\right)^{1/4}.
\label{eq:dfd_min_both}
\end{equation}
For $\epsilon_{\mathrm T}=\epsilon_{\mathrm{L}}=0.1$, and $\Delta f_{\mathrm{L}}=\SI{1.3}{\mega\hertz}$, Eq.~(\ref{eq:dfd_min_both}) gives
\begin{align}
\Delta f_{\mathrm D,\min}
&\simeq
\Delta f_{\mathrm{L}}\left(\epsilon_{\mathrm T}^2+\epsilon_{\mathrm{L}}^2\right)^{1/4}
=
\SI{1.3}{\mega\hertz}\,\left(0.1^2+0.1^2\right)^{1/4}
\nonumber\\
&=
\SI{1.3}{\mega\hertz}\,(0.02)^{1/4}
\approx
\SI{0.49}{\mega\hertz}.
\label{eq:dfd_min_numeric}
\end{align}
This corresponds to a minimum measurable temperature of around \SI{68}{\micro\kelvin}.

For a case in which the laser-frequency jitter contributes only
\SI{100}{\kilo\hertz} to the linewidth, we estimate a minimum detectable
Doppler broadening of
\(\Delta f_{\mathrm D,\min} \simeq \SI{38}{\kilo\hertz}\), corresponding to
a minimum measurable temperature of approximately \SI{411}{\nano\kelvin}.

In the extreme limit where the laser-frequency jitter is much smaller than
the intrinsic linewidth of the transition, the minimum detectable Doppler
broadening is set by the intrinsic linewidth itself. For the
\(\mathrm{70S}_{1/2}\) Rydberg state, which has an intrinsic linewidth of
\(\SI{0.95}{\kilo\hertz}\), this gives a minimum detectable Doppler
broadening of \(\SI{357}{\hertz}\), corresponding to a minimum measurable
temperature of approximately \SI{36}{\pico\kelvin}.


\end{document}